\documentclass[review,number,sort&compress,3p,times]{elsarticle}	

 \usepackage{lineno}
 \usepackage{color}

\usepackage{color}
\usepackage{amssymb}
\usepackage{subfigure}
\begin{document}
\begin{frontmatter}
\title{Effect of humid air exposure on photoemissive and structural properties of KBr thin film photocathode}

\author{R. Rai, Triloki, N. Ghosh}
\author {B. K.~Singh\corref{cor}}
\cortext[cor]{Corresponding author}
\ead{bksingh@bhu.ac.in}
\address{High Energy Physics laboratory, Physics Department,
  Banaras Hindu University,
  Varanasi-221005, India}

\begin{abstract}
  
  We have investigated the influence of water molecule absorption on photoemissive and structural properties of potassium bromide (KBr) thin film photocathode under humid air exposure at relative humidity 62 $\pm$~3$\%$. It is evident from photoemission measurement that the photoelectron yield of KBr photocathode is degraded exponentially with humid air exposed time. Structural studies of the ``as-deposited'' and ``humid air aged'' films reveal that there is no effect of relative humidity on film's crystalline face centered cubic structure. However, the crystallite size of ``humid air aged'' KBr film has been increased as compared to ``as-deposited'' one. In addition, topographical properties of KBr film  are also examined by means of scanning electron microscope, transmission electron microscope and atomic force microscope and it is observed that granular characteristic of film has been altered, even for short exposure to humid atmosphere.
 \end{abstract}

\end{frontmatter}

\section{Introduction}   
 Alkali halide (A-H) thin film photocathodes are incredibly used as a photon converter in the extreme ultraviolet (EUV, 10 nm \textless $ \lambda$ \textless 100 nm) and far ultraviolet (FUV, 100 nm \textless $\lambda$ \textless 200 nm) spectral ranges. These photocathodes are persistently employed in astroparticle detectors, vacuum and gas-based photon detectors, in the detection of scintillation light, medical imaging, in a positron emission tomography~\cite{Triloki} etc. A-H photocathodes are also served as a protective layer in visible-sensitive photon detectors~\cite{A. Breskin1996}. In particular, KBr photocthode is a potential alternative in the astrophysics experiments such as FUSE (The Far Ultraviolet Explorer)~\cite{over}, the EUV spectrometer SUMER ~\cite{SUMER} and an ultraviolet spectrometer, PHEBUS (Probing of Hermean Exosphere by Ultraviolet Spectroscopy)~\cite{PHEBUS} due to sensitivity in the FUV region only (\textless 160 nm ). Owing to sensitivity in FUV region, it  improves the ability to reject sources of radiation and background near ultraviolet (UV) wavelength. These properties of KBr are also advantageous for soft X-ray instruments~\cite{Oswald}. 

The stability and photoelectron yield are vital for determining the performance of photon detectors, where long operation time and high quantum yield are required. Due to hygroscopicity of A-H materials, the quantum yield of photocathode is degraded, even for nominal exposure to humid atmosphere, during the transfer and mounting of the sample. The condensation of water molecules on the A-H thin film surfaces also affect the topographical and structural characteristics of a photocathode. Although enormous statistics are available from previous studies on humid ripening of cesium iodide and other A-H photocathodes~\cite{Yuguang,Braem,Singh,Hoedlmoser,Nitti,Mauro,Singh1,triloki}, very less work has been done on the ageing of KBr photocathode under impact of water molecule absorption. Therefore, in this manuscript, we report on a detailed study aimed at elucidating  the effect of humid air exposure on KBr thin film photocathode.

\section{Experimental techniques}

KBr film of 300 nm thickness was deposited on aluminum (Al) discs and formvar coated copper grids by the thermal evaporation technique, in a high vacuum environment ($4\times10^{-7}$ torr). Water vapor and other residual contaminants inside the chamber were monitored by a residual gas analyzer (SRS RGA 300) before the sample evaporation. KBr powder of 5N purity (Alfa Aesar) was sublimated at the rate of $\leq$ 2 nm/s from a tantalum boat. Rate of deposition and thickness of the films were monitored by a quartz crystal thickness/rate monitor (Sycon STM-100). Prior to deposition, the boat and powders were throughly outgassed up to the temperature close to the melting point of KBr under a shutter. The distance between evaporation source and substrate was kept about 20 cm. Immediately after film preparation, the chamber was purged with dry N$_{2}$, in order to ensure the minimum contact with atmospheric air during the sample transfer and films were extracted into a vacuum desiccator, containing fresh silica gel. Further, KBr films were transported for photoemission, X-ray diffraction (XRD), scanning electron microscopy (SEM), transmission electron microscopy (TEM) and atomic force microscopy (AFM) measurements.

Photoemissive properties were evaluated by measuring the photocurrent from KBr films under UV illumination. VUV monochromator (Model: 234/302, Mcpherson), with base pressure $1.73\times10^{-4}$ torr, was exploited to generate a monochromatic light of spectral range of 110 to 160 nm. 
This VUV monochromator was equipped with a 30 W magnesium fluoride (MgF$_{2}$) windowed deuterium (D$_{2}$) lamp. The stability of D$_{2}$ lamp was monitored during the entire experiment by Hamamatsu's made PMT (Model: 658). A positive voltage (+200 V) was applied on a grid (anode) kept at a 2 mm distance from KBr photocathode in order to collect the UV induced photoelectrons and resulting photocurrent was measured by a keithley picoameter (Model: 6485).

 In order to examine crystallographic nature of KBr film, XRD line profile data was recorded in continuous scan mode (2$\theta= 10^{o}
-85^{o}$) using X' Pert PRO $\theta/2\theta$ (PANanalytical) diffractometer in the Bragg-Brentano parafocusing configuration with CuK$\alpha$ ($\lambda$ = 0.15406~\AA) radiation. The diffraction beam optics contains 0.04 rad solar slit and a scintillator detector. To minimize the instrumental contribution in XRD line broadening, diffractometer has been calibrated with standard silicon (Si) crystal. 

 SEM images of KBr film were scanned by FEI Quanta $200$ at 10 kV accelerating voltage with secondary electron detector in $5\times10^{-4}$~torr vacuum environment and TEM micrograph was recorded by TECNAI-20$G^{2}$, operated at 200 kV voltage in the bright field diffraction and imaging modes with single tilt sample stage. Atomic force microscope (AFM) ND-MDT solver-NEXT, coupled with the PX Ultra controller and NOVA PX data processing software was employed to investigate the influence of moisture on surface roughness and average grain height of KBr film. All micropatterns were recorded at 24 $\pm$ 3$~^{o}C$ temperature.

\section{Result and Discussion}
\subsection{Photoemission studies} 
The photoemission properties of 300 nm thick KBr film has been investigated in 110 to 160 nm wavelength range. Initially, photocurrent of ``as-deposited'' KBr film was measured. Then the sample was exposed to humid air (RH = 62 $\pm$~3$\%$) at the interval of two hours each and corresponding photocurrent was recorded. Thereafter, this experiment was consecutively iterated at 1 hour interval (from 4 hours to 9 hours) and influence of moisture on the photocurrent of KBr film was analyzed. During humid air exposure, the water content is more than 20 000 ppmv at 25$~^{o}C$ temperature~\cite{ppmv}. The variation of relative photocurrent with exposed time is shown in Figure 1. Here, we analyzed the effect of humidity on few selected wavelengths (125, 131, 138 and 144 nm respectively) only. From photoemission measurements, it is observed that photoelectron yield has been decreased with exposure time and this reduction is more manifested at higher wavelengths (138 and 144 nm).  

\begin{figure}[!ht]
   \begin{center}
    \includegraphics[scale=0.35]{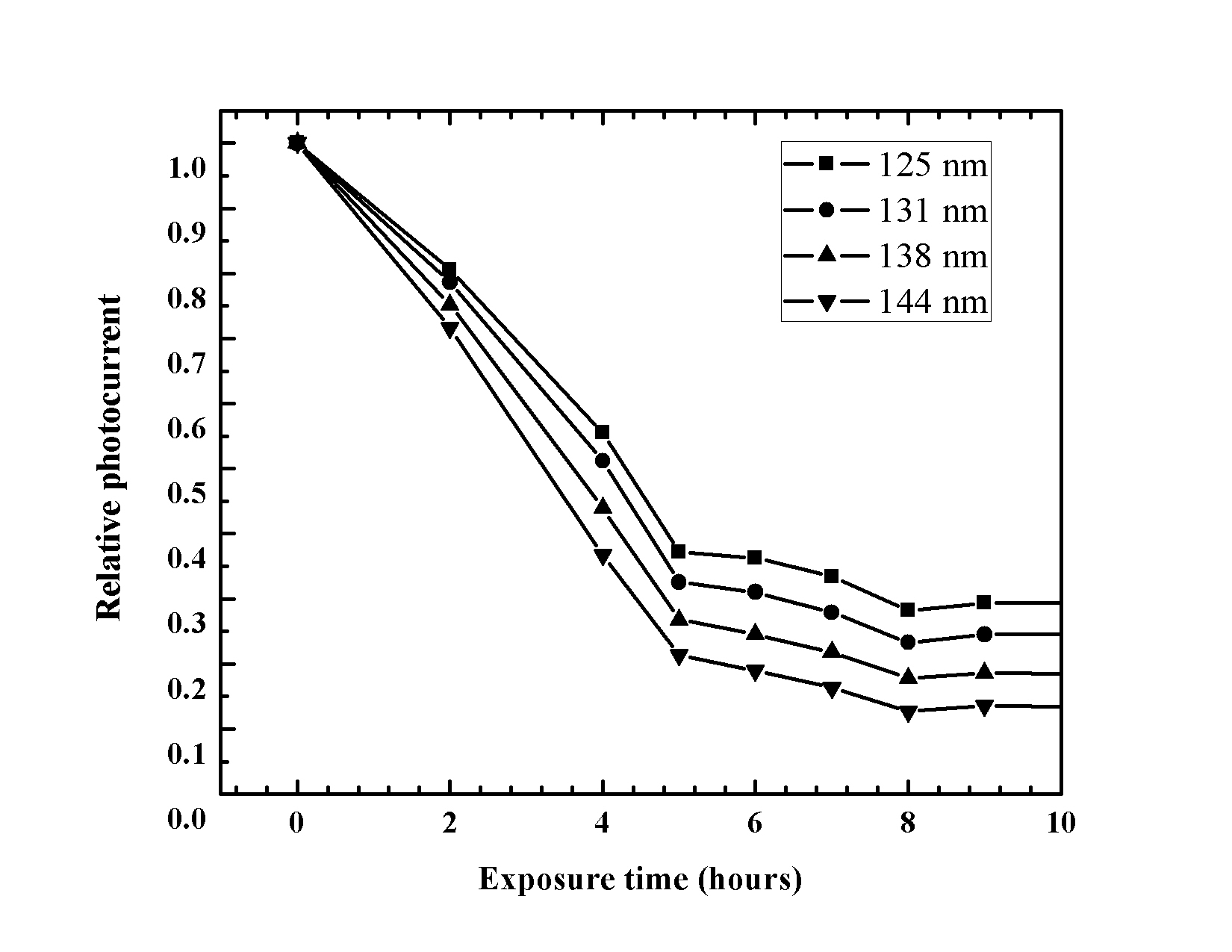}
    \caption{Degradation in relative photocurrent as a function of time with humid air exposure on wavelengths 125, 131, 138 and 144 nm for 300 nm thick KBr film.}
    \label{fig1}
  \end{center}
  \end{figure}
In the degradation of photocurrent, water (H$_{2}$O) molecules play a significant role. Due to unequal distribution of electronic charge pairs and bent shape (H-O-H) of water molecules, it has a polar nature. The O-end acts as a slightly negative pole and H-end acts as a slightly positive pole. When KBr film is exposed to moisture, the oxygen atom (negative pole) is absorbed preferentially on positive ion (K$^{+}$) and H-atoms (positive pole) directed towards negative ion (Br$^{-}$) side as shown in Figure 2 . For simplicity reason, we are showing a flat surface (Figure 2). But this is not true in practice, since kink or step has been appeared on the surface of thermally deposited film. 

 As the humidity increases with time, water molecules will congregate over the orderly array of ions at granular's surface and will form probably a hydrogen bond with each other. This leads to formation of water cluster or chain on KBr film surface, in which KBr dissolves and will give rise to the increment in the ionic mobility of KBr film with increase in number of dissolved ions~\cite{Luna}. When we applied +200 V on anode (grid), anion (Br$^{-}$ ion) will attract towards the positive potential and shift the surface potential of film at negative value, which affect the electron transport properties of KBr photocathode.  The degradation in photocurrent is also more pronounced due to higher electron affinity (3.36 eV) of Br$^{-}$ ion .

\begin{figure}[!ht]
  \begin{center}
    \includegraphics[scale=0.4]{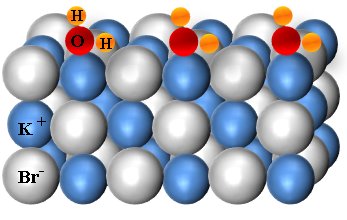}
    \caption{Directionality of O-H bonds along (100) plane of water molecule absorbed on KBr thin film photocathode surfaces.}
   \label{fig2}
  \end{center}
  \end{figure}

\subsection{X-ray diffraction analysis}
X-ray line profile analysis is an ideal tool to quantify the effect of humid air exposure on microstructural parameters of KBr films. XRD patterns of ``as-deposited'' and ``humid air aged'' films, evaporated on Al disc, are shown in Figure 3. The XRD profile exhibits a most intense peak on diffraction angle $2\theta$ = 27.1453$^{o}$, followed by $2\theta$ = 55.7677$^{o}$ and 82.4375$^{o}$. First peak corresponds to (200) crystallographic plane and other two to (400) and (440) planes respectively, which attributes to crystalline face centered structure of KBr film.  These peak positions are matched with ASTM card data (pdf no.-730831). The lattice constant ($a$) is calculated by an analytical relation: 

\begin{equation}
a=d\sqrt{h^{2} + k^{2} + l^{2}}
\end{equation}
where, $d$ is interplaner separation between adjacent planes and ($hkl$) are the index of the diffracted plane. The value of lattice constant corresponding (200) plane is obtained about 6.57016~$\AA$ and 6.56372~\AA~respectively for ``as-deposited'' and ``humid air aged'' (RH = 65\%)
KBr film.

To examine the effect of humidity on the length of coherent scattering domain i.e. crystallite size, the well known Scherrer relation is employed 
~\cite{Scherrer}:

\begin{figure}
\begin{center}
\includegraphics[width=80mm]{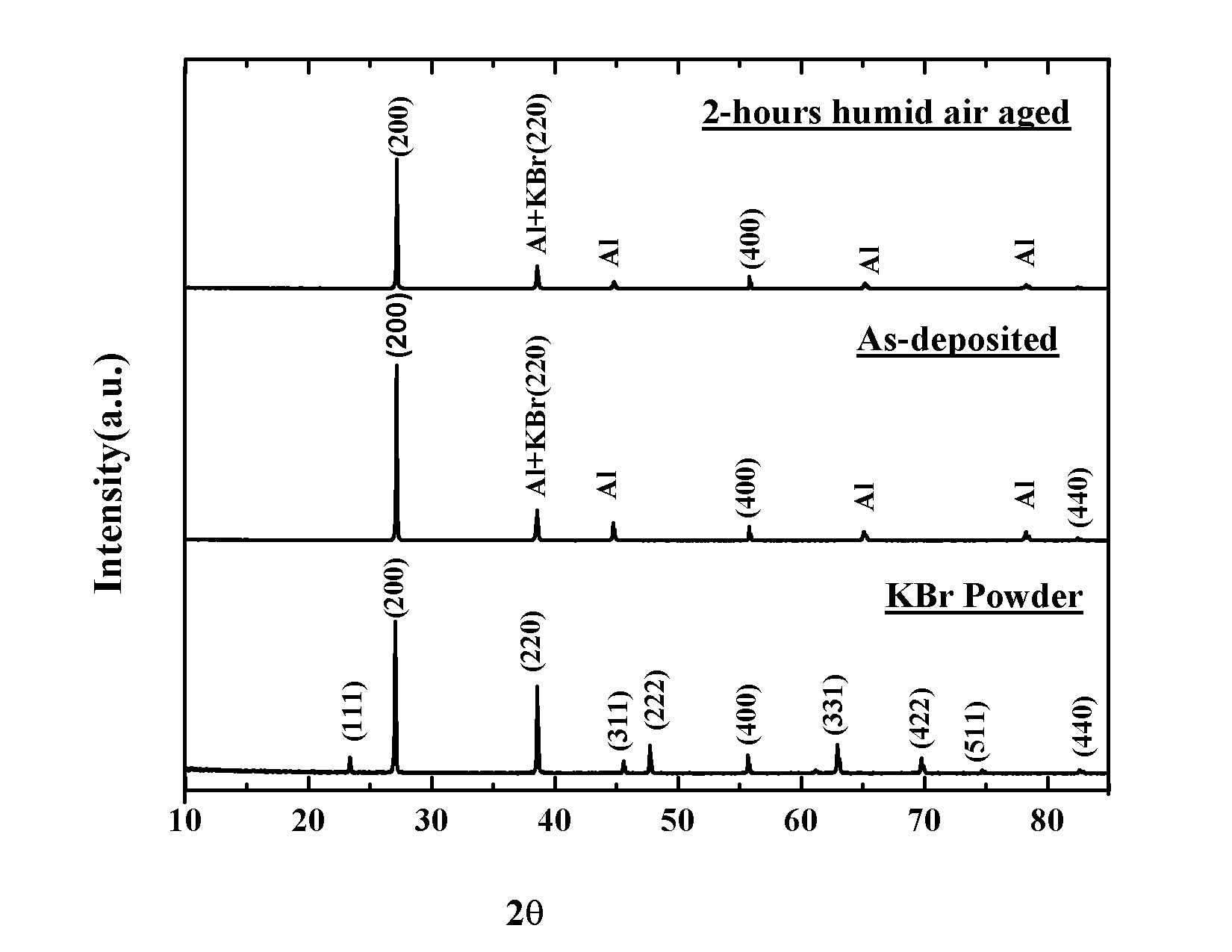}
\caption{\label{fig3} XRD pattern of ``as-deposited'' and ``2-hours humid air aged'' film and of KBr powder.}
\end{center} 
\end{figure}

\begin{equation}
  D= \frac{K \lambda }{ \beta_{hkl} \cos\theta} 
\end{equation}
where, $D$ is the effective crystallite size normal to the reflecting plane (200), $K$ is a shape factor (0.9), $\lambda$ is the wavelength of CuK$\alpha$ radiation, $\beta$$_{hkl}$ is full width at half maximum (FWHM) of particular peak and $\theta$ is the diffraction angle. The estimated value of crystallite size is found to be about 68 nm in the case of ``as-deposited'' film and 82 nm after 2 hours humid air exposure. It is evident from the determination of microstructural parameters (lattice constant and crystallite size) of 300 nm KBr film that inter planner spacing between two adjacent crystallographic planes as well as the linear dimension of coherently scattering domain have been increased with humid air exposure.

\subsection{SEM analysis}
  SEM images of ~``as-deposited'' and ``2-hours humid air aged'' (RH = 65 $\pm~5\%$) films are shown in Figure 4. We observed that the average grain size is increased, while total area coverage and grain density have been decreased with humid air exposure. The average grain size is 566 nm in the case of ``as-deposited'' KBr film, while 757 nm for aged film. The extension in grains, after exposure to humid air is arisen due to coalescence process. Prior to coalescence, there was a collection of various grains of different sizes. With humid air exposure of KBr film, the larger grains  grow at the expense of the smaller ones. The time evolution of the distribution of grain sizes can be derived by a desire to minimize the surface free energy  through a decrease in surface area of the grain structure. Absorption of water molecules on surface curvature of grains gives rise to local concentration differences that are alleviated by mass flow from smaller grain (smaller surface area) to larger one. Several mechanisms are available for mass transport, two most likely ones involve self diffusion through a bulk or via the surfaces of the grains. Diffusion will proceed from the smaller grain to larger grain until the former disappears entirely~\cite{book}.

 \begin{figure}
\begin{center}
\includegraphics[height=30mm]{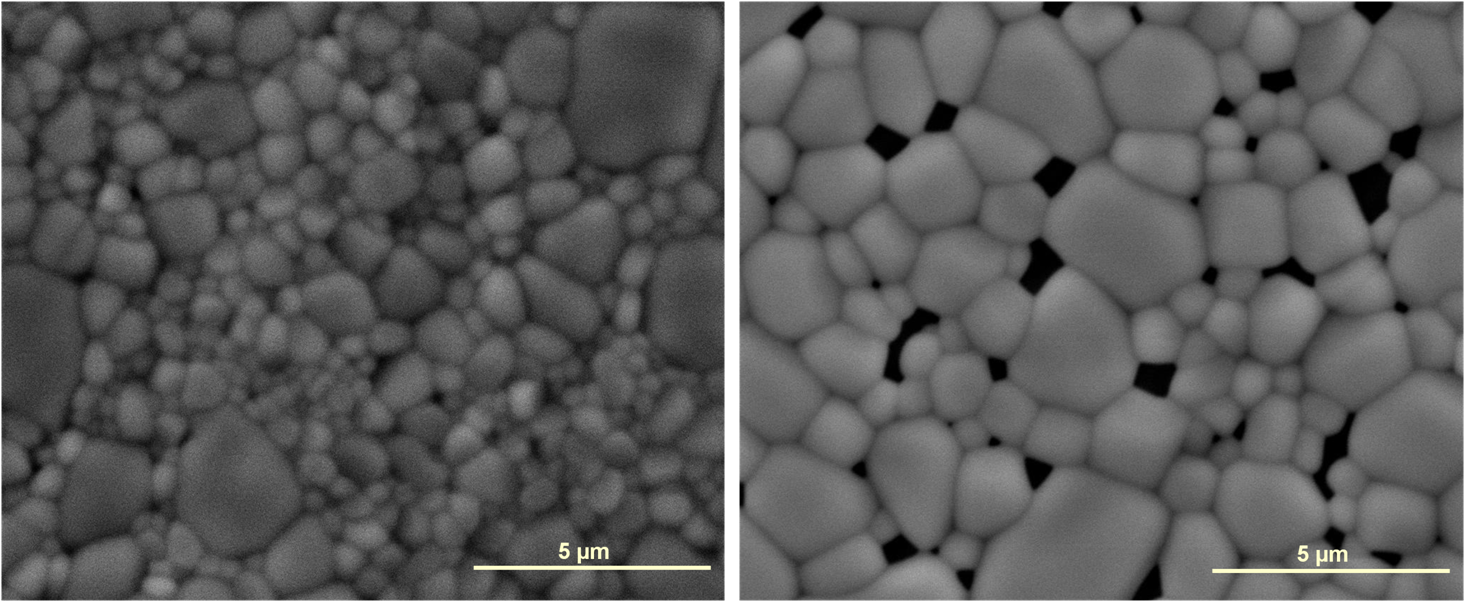}
\caption{\label{fig4} Scanning electron microscopy image of ``as-deposited'' (left panel) and ``2-hours humid air aged'' (right panel) of 300 nm KBr  film deposited on Al disc.} 
\end{center}
\end{figure}

\subsection{AFM analysis}
Surface morphology obtained by AFM analysis of 300 nm ``as-deposited'' and ``2-hours humid aged'' KBr film is shown in Figure 5. In AFM, a sharp tip (V-shape) at the end of cantilever was scanned over a surface. While scanning, surface features deflect the tip and thus cantilever. By measuring the deflection of the cantilever, a topographic image of the film surface is acquired ~\cite{AFM}. Here, we used 5 $\mu$m $\times$ 5 $\mu$m topological scale to investigate the surface morphology of 300 nm thick KBr film because it has better resolution as compared to higher order scale (10 $\mu$m or 20 $\mu$m). In the case of ~``as-deposited'' film, surface appears to consist multifaceted grain with more than 90$\%$ substrate coverage. After 2-hours humid air exposure (RH = 65\%) at~25 $^{o}C$ temperature, the surface continuity as well as grain size is affected. The average grain size has been increased from 332 to 477 nm for humid aged film. Further, maximum peak to valley height $Z_{max}$ decreases, however average height $Z_{avg}$ increases with water molecules absorption (see table 1). It is also observed that root mean square roughness, defined as the standard deviation of the elevation, within the given area (5 $\mu$m $\times$ 5 $\mu$m) has been increased after exposure to humid atmosphere. But, these deviation in $Z$ parameter are very small (order of few nm), therefore we can not make any conclusion. For analyzing the effect of water molecule absorption on peak height, we have to expose the sample for few more hours to humid atmosphere.

\begin{figure}
\begin{center}
\includegraphics[height=70mm]{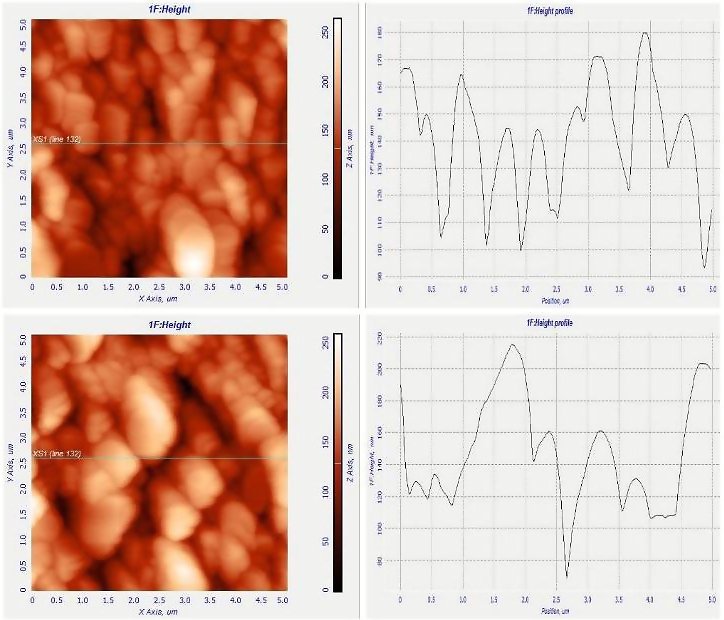}
\caption{\label{fig4} 2D AFM images and representative peak height ($Z$) as a function of distance of  ``as-deposited'' (top) and ``2-hours humid air aged'' (bottom) of 300 nm KBr  film deposited on Al disc.}
\end{center}
\end{figure}

\begin{table*}[tp]
\renewcommand{\arraystretch}{1.2}
\caption{AFM profile of ``as-deposited'' and ``2~hours humid air aged'' KBr film}
 \vskip .1cm

\centering

\begin{tabular}{c  c  c  c  c}

\hline

300 nm thick & Average grain & Root mean square          & Maximum Peak to        &   Average    \\ 
KBr film     & size (nm)     & roughness, $R_{rms}$ (nm)  & valley  height $Z_{max}$ (nm) &   height $Z_{avg}(nm)$ \\ [1.0ex]

\hline

As-deposited & 333 & 33.12 & 262 & 141 \\

Humid aged & 477 & 37.14 & 230 & 147  \\

\hline

\end{tabular}
\label{table:nonlin}

\end{table*}

 \subsection{TEM analysis}
 TEM micropattern  endeavors to drive a topographical and structural features of the sample. To ensure that the observed grain morphology and crystal structure (fcc) is representative of entire film surface, few regions of the KBr film are scanned by TEM, but here we report only one bright field image in Figure 6. For the fresh sample, the film surface appears to be continuous with interconnected grains. After exposing to the humid atmosphere (RH = 62\%) with water contain  19863.0 ppmv,  a crack or discontinuity appears in grain distribution as well as neighboring grains are diffused and form a bigger one in order to minimize the surface area. The average grain size is about 145 nm for ``as-deposited'' film and 286 nm for ``humid air aged'' film. It is also visible from TEM micrograph that ageing starts from the grain boundaries.    

The selected area diffraction pattern (SAED) shows that film has polycrystalline structure (fcc) as  bright diffraction spots occur at all possible azimuthal angles and  appear to be originate on the perimeter of the circle. The value of lattice constant $(a)$ was manually calculated for 150 mm camera length (camera constant = 4.8466 cm\AA) and it found to be about 6.3284~\AA~for ``as-deposited'' film, which have small difference from of ASTM card data ($a$ = 6.5847~\AA). This difference may be arisen due to finite crystallite size and lattice strain. However in the case of ``humid air aged'' film, lattice constant is 6.4897~\AA, which shows that KBr film is more relax and forms large crystallite after water vapor absorption. This result is also boosted by crystallite size obtained by a XRD measurement.

 \begin{figure}
\begin{center}
\includegraphics[height=30mm]{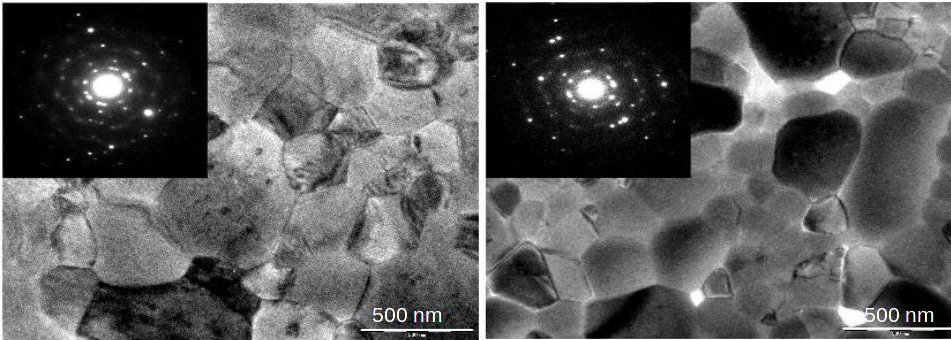}
\caption{\label{fig4} Transmission electron microscope (TEM) images and SAED pattern of 300 nm ``as-deposited'' (left) and ``2-hours humid air aged'' (right) of KBr film.}
\end{center}
\end{figure}

 \section{Conclusion}
 Degradation in relative photoelectron yield is observed and it is found that this reduction in photocurrent as a function of exposed time is not linear, but probably exponential. XRD and TEM SAED diffraction pattern reveals that KBr film has purely crystalline in nature. Most intense XRD peak is obtained on (200) crystallographic plane, followed by two other peaks on (400) and (440) plane, indicating a crystalline face centered cubic structure. Crystallite size (calculated using Scherrer's method) has been increased after exposure to humid atmosphere. Grain like morphology is evident from TEM, SEM and AFM micropattern, with more that 90\% surface area coverage and average grain size is about 286 nm, 566 nm and 333 nm respectively for ``as-deposited'' film, which has been increased after exposure to humid atmosphere. Discrepancy in average grain size, obtained from TEM, SEM and AFM measurement is appeared due to variation in resolution of these three morphological tools.  In  AFM analysis, we found the growth in the average roughness and average height of the grains, while maximum peak to valley height~$Z_{max}$ has been decreased after humid ageing.   
      
 \section{Acknowledgment}
  This work was partially supported by the Department of Science and Technology (DST), the Council of Scientific and Industrial Research (CSIR) and the Indian Space Research Organization (ISRO). Triloki and R. Rai acknowledge University Grant Commission (UGC), New Delhi, India for providing financial support.

\end{document}